\documentclass[12pt]{article}
\usepackage{graphicx}
\begin{document}

\begin{center}

{\Large \bf
Internal Space-time Symmetries according to Einstein, Wigner, Dirac, and Feynman}

\vspace{2ex}

Y. S. Kim~\footnote{email: yskim@physics.umd.edu}\\
Center for Fundamental Physics, University of Maryland, \\
College Park, Maryland 20742, U.S.A.
\vspace{2ex}

Marilyn E. Noz~\footnote{email: marilyne.noz@gmail.com} \\
Department of Radiology, New York University,\\
New York, New York 10016, U.S.A.

\end{center}

\vspace{2ex}

\begin{abstract}
When Einstein formulated his special relativity in 1905, he
established the law of Lorentz transformations for point particles.
It is now known that particles have internal space-time structures.
Particles, such as photons and electrons, have
spin variables.  Protons and other hadrons are
regarded as bound states of more fundamental particles called
quarks which have their internal variables.  It is still one of the most
outstanding problems whether these internal space-time variables
are transformed according to Einstein's law of Lorentz transformations.
It is noted that Wigner, Dirac, and Feynman made important
contributions to this problem.  By integrating their efforts, it
is then shown possible to construct a picture of the internal
space-time symmetry consistent with Einstein's Lorentz covariance.
\end{abstract}

\newpage
\section{Introduction}\label{intro}
It took Issac Newton twenty years to extend his law of gravity from
two particles to two extended objects, such as the sun and the
earth. To do this, he had to develop a new mathematics now called
integral calculus.
\par
When Einstein formulated his special relativity in 1905, his
transformation law was for point particles.  We still do not know
what happens to classical rigid bodies, but quantum mechanics allows
us to replace those by standing waves.  In the case of hydrogen atom,
one can argue that the circular orbit appears as an ellipse to a
moving observer~\cite{bell04}, but not much has been done beyond this.
\par
In quantum mechanics, the hydrogen atom is a localized probability
distribution constructed from a standing wave solution of Schr\"odinger's
wave equation.  We do not observe too often hydrogen atoms moving with
relativistic speed, but relativistic-speed protons are abundantly
produced from accelerators.  Like the hydrogen atom, the proton is
a bound state.  The constituents are the quarks.
\par
In 1939~\cite{wig39}, Eugene Wigner published a paper on the little
groups of the Lorentz group whose transformations leave the
four-momentum of a given particle invariant.  He showed that the
little groups for particles with non-zero mass are isomorphic to the
three-dimensional rotation group.  If the particle is at rest, this
symmetry group is the three-dimensional rotation group.  In this way,
Wigner was able to define the particle's spin as a space-time variable
in the Lorentz covariant world.  In 1939, composite particles, such as
hadrons in the quark model, were unthinkable.
\par
In a series of papers from 1927 to 1963, Paul A. M.  Dirac attempted
to construct a Lorentz-covariant picture of localized wave functions.
In 1927~\cite{dir27}, he produced the concept of the c-number
time-energy uncertainty relation and noted the basic space-time
asymmetry.  In 1945~\cite{dir45}, he considered the possibility of
harmonic oscillators which can be Lorentz-transformed.  In
1949~\cite{dir49}, he formulated the technique of light-cone variables
to deal with Lorentz boosts.  In 1963~\cite{dir63}, Dirac observed
that two coupled oscillators can produce the symmetry of the Lorentz
group of three space coordinates and two time variables.
\par
In 1971~\cite{fkr71}, Feynman {\em et al.} attempted to understand
the hadronic mass spectra in terms of the degeneracies of the
three-dimensional harmonic oscillator.  In so doing they wrote down
a harmonic oscillator differential equation which takes the same
form for all Lorentz frames.  This Lorentz-invariant differential
equation has many different solutions, but they chose a set of
solutions that violate all the rules of quantum mechanics and
relativity.  We show in this paper that their equation can have
solutions which are consistent with the observations made earlier
by both Wigner and Dirac.
\par
In 1969~\cite{fey69a,fey69b}, Feynman had observed that the
ultra-fast proton can appear like a collections of partons while
the proton is like a bound state of quarks.  Since the partons
have properties quite different those of the quarks, Feynman's
parton picture presents a nontrivial covariance problem.  In 1905,
Einstein had the problem of showing that the energy-momentum
relation takes different forms for slow and fast particles.
\par
In this paper, we review the efforts made by Wigner, Dirac, and
Feynman in Secs.~\ref{wigner}, \ref{dirac}, and \ref{feynman}
respectively.  We then integrate their contributions in
Sec.~\ref{kimnoz} to produce a Lorent-covariant picture of
quantum bound states.  Finally, in Sec.~\ref{quarkmo}, we
discuss experimental consequences of this covariant formalism.

\section{Wigner's Little Groups}\label{wigner}
In his 1939~\cite{wig39}, Wigner constructed subgroups of the
Lorentz group whose transformations leave the four-momentum of
a given particle invariant.  They are called the little groups.
The little groups are isomorphic to the $O(3)$ and to $O(2,1)$
groups if the particle momentum is time-like and space-like
respectively.  If the four-momentum is light-like, the little
group is isomorphic to the two-dimensional Euclidean group.
Since the momentum remains invariant, the little groups dictate
the internal space-time symmetries of particles in the
Lorentz-covariant world.
\par

Since it is well known that the $SL(2,c)$ group serves as the
covering group of the Lorentz group, it is possible to explain
Wigner's little groups in terms of two-by-two matrices.  Let us
consider the unimodular matrix
\begin{equation}
\pmatrix{A & B \cr C & D}
\end{equation}
where all four elements are real numbers with $(AD - BC) = 1.$
There are thus three independent parameters.  This matrix can then
be rotated to one of the following equi-diagonal matrices.
\begin{equation}\label{wigmat}
 \pmatrix{\cos\theta & -e^{-\eta}\sin\theta \cr
       e^{\eta}\sin\theta & \cos\theta},
\qquad
 \pmatrix{\cosh\lambda & e^{-\eta}\sinh\lambda \cr
           e^{\eta}\sinh\lambda & \cosh\lambda},
\qquad
\pmatrix{1 & 0 \cr \gamma & 1} .
\end{equation}


This is purely a mathematical statement.  However, they form
the basis for Wigner's little groups for massive particles,
imaginary-mass particles, and massless particles
respectively~\cite{kim10}.
\par
Let us look at the first matrix in Eq.(\ref{wigmat}).  It
can be written as
\begin{equation}
\pmatrix{ e^{-\eta/2} & 0 \cr 0 & e^{\eta/2}}
\pmatrix{\cos\theta & \sin\theta \cr \sin\theta & \cos\theta}
\pmatrix{ e^{\eta/2} & 0 \cr 0 & e^{-\eta/2}} ,
\end{equation}
which corresponds to a Lorentz boost of the rotation matrix along the $z$
direction.  The rotation matrix performs a rotation around the $y$ axis.
The little group for the massive particle is a Lorentz-boosted rotation
group.
\par
Wigner noted in 1939 that, for a massive particle, there is a Lorentz
frame where the particle is at rest.  In this frame, rotations leave the
particle four-momentum invariant, it can rotate internal space-time variables.
The particle spin is the prime example.  Wigner noted further that
the third matrix of Eq.(\ref{wigmat}) corresponds to the little group for
massless particles.  Then there are two questions.  The first question is
what physical variable does $\gamma$ correspond to?  The second question
is whether this triangular matrix is a limiting case of the first matrix.
\par
\begin{figure}[thb]
\centerline{\includegraphics[scale=0.4]{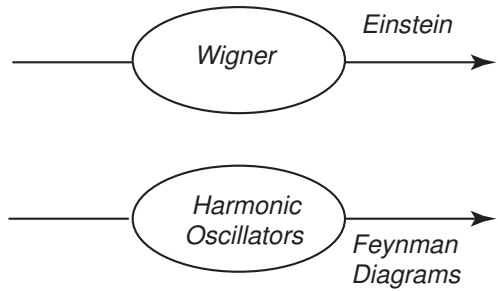}}
\caption{According to Einstein, point particles obey the Lorentz
transformation law.  We expect that the extended particles should
obey the same law.  Transformations of Wigner's little groups
leave the external momentum invariant, but change the internal
space-time variables.  In quantum field theory, Feynman diagrams
describe running waves according to Einstein's Lorentz covariance.
It is possible to construct an oscillator-based model for
internal space-time structure consistent with Wigner and thus
with Einstein.}\label{dff77}
\end{figure}

Let us answer the second question first.  If the particle mass approaches
zero, the $\eta$ parameter becomes infinitely large.  If it is allowed to
become large with $\theta e^{\eta/2} = \gamma,$ the angle $\theta$ has to
become zero with $\cos\theta = 1.$  The variable $\gamma$ is for the
gauge transformation.  These answers have a stormy history, but a
geometric picture was developed by Kim and Wigner in 1990~\cite{kiwi90jmp}.
Wigner's little group is compared with Einstein's energy-momentum relation
in Table~I.
\vspace{3mm}
\begin{quote}Table I. Covariance of the energy-momentum relation, and
covariance of the internal space-time symmetry groups.\end{quote}
\begin{center}
\begin{tabular}{ccc}
Massive, Slow & COVARIANCE & Massless, Fast \\[2mm]\hline
{}&{}&{}\\
$E = p^{2}/2m$ & Einstein's $E = mc^{2}$ & $E = cp$ \\[4mm]\hline
{}&{}&{}  \\
$S_{3}$ & {}  &    $S_{3}$ \\ [-1mm]
{} & Wigner's Little Group & {} \\[-1mm]
$S_{1}, S_{2}$ & {} & Gauge Trans. \\[4mm]\hline
\end{tabular}
\end{center}
\vspace{3mm}
In 1939, it was unthinkable that the proton is a composite particle
and is a bound state of the quarks.  It can also move with a speed
very close to that of light.  In Sec.~\ref{feynman}, we shall see
whether this quantum bound state has the symmetry of Wigner's little
group using harmonic oscillators.  This plan is illustrated in
Fig.~\ref{dff77}.
\vspace{3mm}
\vspace{1mm}
\begin{figure}[thb]
\centerline{\includegraphics[scale=0.42]{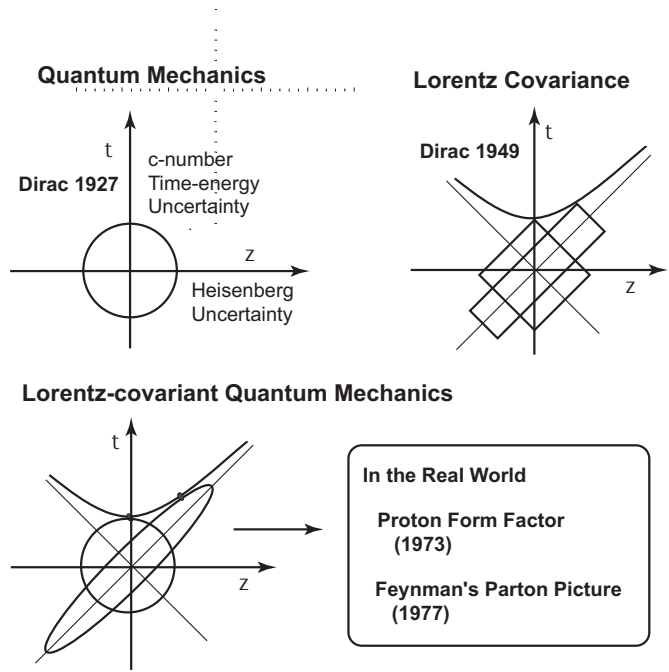}}
\caption{Space-time picture of quantum mechanics.  In his 1927 paper,
Dirac noted that there is a c-number time-energy uncertainty
relation, in addition to Heisenberg's position-momentum uncertainty
relations, with quantum excitations.  This idea is illustrated
in the first figure (upper left).  In his 1949 paper, Dirac produced
his light-cone coordinate system as illustrated in the second figure
(upper right). It is then not difficult to produce the third figure,
for a Lorentz-covariant picture of quantum mechanics.}\label{dirkn}
\end{figure}
\vspace{1mm}

\section{Dirac's Attempts to make Quantum \\Mechanics Lorentz
Covariant}\label{dirac}
Paul A. M. Dirac made it his lifelong effort to make quantum
mechanics consistent with special relativity.  In 1927~\cite{dir27},
Dirac notes that there is an uncertainty relation between the
time and energy variables which manifests itself in emission of
photons from atoms.   He notes further that there are no excitations
along the time or energy axis, unlike Heisenberg's uncertainty
relation which allows quantum excitations.  Thus, there is a serious
difficulty in combining these relations in the Lorentz-covariant world.
\par
In 1945~\cite{dir45}, Dirac considered a four-dimensional harmonic
oscillator and attempted to construct a representation of the Lorentz
group using the oscillator wave functions.  However, he ends up with
wave functions which do not appear to be Lorentz-covariant.
\par
In 1949~\cite{dir49}, Dirac considered three forms of relativistic
dynamics which can be constructed from the ten generators of the
Poincar\'e group.  He then imposed subsidiary conditions necessitated
by the existing form of quantum mechanics.  In so doing, he ends up
with inconsistencies in all three of the cases he considers.  On the
other hand, he introduced the light-cone coordinate system which
allows us to perform Lorentz boosts as squeeze
transformations~\cite{kn73}
\par
In 1963~\cite{dir63}, he constructed a representation of the $O(3,2)$
deSitter group using two harmonic oscillators.  Using step-up and
step-down operators, he constructs a beautiful algebra, but he made
no attempt to exploit the physical contents of his algebra.  Indeed,
his representation now serves as the fundamental scientific language
for squeezed states of light, further enforcing the point that
Lorentz boosts are squeeze transformations~\cite{knp91}.
\par
We can combine Dirac's time-energy uncertainty relations and his
light-cone coordinate system to obtain a Lorentz-covariant picture
of quantum mechanics, as shown in Fig.~\ref{dirkn}.

\section{Feynman's Phenomenological Equation for both Scattering
and Bound States}\label{feynman}
In order to explain the hadron as a bound state of quarks,
Feynman {\em et al.} start with two quarks whose space-time
coordinates are
$x^a_{\mu}$ and $x^b_{\mu}$ respectively~\cite{fkr71} by using
the equation
\begin{equation}\label{diff11}
\left\{-\frac{1}{2}\left[\left(\frac{\partial}
                         {\partial x^a_{\mu}}\right)^2 +
\left(\frac{\partial}{\partial x^b_{\mu}}\right)^2 \right]  +
\frac{1}{16}\left(x^a_{\mu} -x^b_{\mu}\right)^2 + m_0^2\right\}
\phi\left(x^a_{\mu}, x^b_{\mu}\right) = 0 .
\end{equation}
If we use the hadronic and the quark separation coordinates as
\begin{equation}\label{coord}
X_\mu = \frac{1}{2}\left(x^a_{\mu} + x^b_{\mu} \right), \qquad
x_\mu = \frac{1}{2\sqrt{2}}\left(x^a_{\mu} - x^b_{\mu}\right) ,
\end{equation}
respectively, it is possible to consider the separation of the
variables:
\begin{equation}
\phi\left(x^a_{\mu},x^b_{\mu}\right) = f\left(X_{\mu}\right)
\psi\left(x_{\mu}\right) .
\end{equation}
Then the differential equation can be separated into the following
two equations.
\begin{equation}\label{diff22}
\left\{\left(\frac{\partial}{\partial X_{\mu}}\right)^2 + m_0^2 +
(\lambda + 1) \right\}
f\left(X_{\mu}\right) = 0
\end{equation}
for the hadronic coordinate, and
\begin{equation}\label{diff33}
\frac{1}{2} \left\{-\left(\frac{\partial}{\partial x_{\mu}}\right)^2
+ x_{\mu}^2 \right\} \psi\left(x_{\mu}\right) = (\lambda + 1)
\psi \left(x_{\mu}\right) ,
\end{equation}
for the coordinate of quarks inside the hadron.
\par
The differential equation of Eq.(\ref{diff22}) is a Klein-Gordon
equation for the Lorentz-invariant hadronic coordinate.  The
differential equation of Eq.(\ref{diff11}) contains the
scattering-state equation for the hadron, and the bound-state
equation for the quarks inside the hadron. Additionally, it is
Lorentz-invariant.
\par
However, in their paper~\cite{fkr71}, Feynman {\it et al.} did not
consider whether their solutions are consistent with the symmetry
of Wigner's little group explained in Sec.~\ref{wigner} of the
present paper.  Let us now construct a representation of Wigner's
little group using the oscillator solutions.  As noted
earlier~\cite{kno79,knp86}, a set of solutions for the oscillator
equation of Eq.(\ref{diff33}) corresponds to a representation of
Wigner's $O(3)$-like little group for massive particles.  If the
hadron is at rest, its wave function should satisfy the $O(3)$
symmetry.  We can achieve this goal by keeping the time-like
oscillation in its ground state, and construct an $O(3)$-symmetric
spatial wave function using the spherical coordinate system.  We
can then write the solution as
\begin{equation}\label{sol00}
\psi(x,y,z,t) = \left[\left(\frac{1}{\pi}\right)^{1/4}
\exp{\left(\frac{-t^2}{2}\right)}\right]\psi(x,y,z) ,
\end{equation}
where the form of $\psi(x,y,z)$ in the spherical coordinate system
is well known.  This spherical solution can also be written as a
linear combination of solutions in the Cartesian coordinate system,
which take the form
\begin{equation}\label{sol33}
\left[\frac{1}{\pi\sqrt{\pi} 2^{(a+b+n)}a!b!n!}\right]^{1/2}
    H_a(x) H_b(y) H_n(z)
\exp{\left\{-\left(\frac{x^2 + y^2 +z^2}{2}\right)\right\}} ,
\end{equation}
where $H_n(z)$ is the Hermite polynomial.
\par
It is now possible to boost this solution along the z direction.
Since the transverse $x$ and $y$ coordinates are not affected by
this boost, we can separate out these variables in the oscillator
differential equation of Eq.(\ref{diff33}), and consider the
differential equation
\begin{equation}\label{diff44}
\frac{1}{2} \left\{\left[-\left(\frac{\partial}{\partial z}\right)^2
+ z^2 \right]
-\left[-\left(\frac{\partial}{\partial t}\right)^2 +
t^2\right]\right\}\psi(z,t) = n \psi(z,t) .
\end{equation}
This differential equation remains invariant under the Lorentz boost
\begin{equation}\label{boostm}
z' = (\cosh\eta) z  -(\sinh\eta) t, \qquad
 t' = (\cosh\eta) t - (\sinh\eta) z,
\end{equation}
where
$$
e^{\eta} = \sqrt{\frac{1 + \beta}{1 - \beta}} $$
with  $\beta = v/c$.
\par
If we suppress the excitations along the $t$ coordinate, the
normalized solution of this differential equation is
\begin{equation}\label{sol44}
\psi(z,t) =  \left(\frac{1}{\pi 2^{n}n!} \right)^{1/2}
H_n(z)\exp{\left\{-\left(\frac{z^2 +t^2}{2}\right)\right\}} .
\end{equation}
We can boost this wave function by replacicing $z$ and $t$ by
$z'$ and $t'$ of Eq.(\ref{boostm}) respectively.

\section{Lorentz-covariant Wave Functions with \\
Physical Interpretation}\label{kimnoz}
While the oscillator equation given in
Eqs.~(\ref{sol44}) and (\ref{sol55}) of Sec.~\ref{feynman} are
solutions of the equation of Feynman {\em et al.},
we should exact meaningful physics from them.
To do this we examine whether it is possible to construct localized
quantum probability distributions.
In terms of the light-cone variables defined in~\cite{dir49},
Eqs.~(\ref{sol44}) and (\ref{sol55}) can be written as
\begin{equation}\label{cwf11}
\psi_{0}
^{n}(x,t) = \left[\frac{1}{\pi n! 2^{n}} \right]^{1/2}
 H_{n}\left( \frac{u + v}{\sqrt{2}}\right)
 \exp{\left\{-\left(\frac{u^{2} + v^{2}}{2}\right)\right\}} ,
\end{equation}
and
\begin{equation}\label{cwf22}
\psi_{\eta}^{n}(x,t) = \left[\frac{1}{\pi n! 2^{n}} \right]^{1/2}
 H_{n}\left(\frac{e^{-\eta}u +  e^{\eta} v}{\sqrt{2}}\right)
\exp{\left\{-\left(\frac{e^{-2\eta}u^{2} + e^{2\eta}v^{2}}{2}
\right)\right\}} ,
\end{equation}
for the rest and moving hadrons respectively.
This form can be expanded as~\cite{knp86}
\begin{equation}\label{cwf33}
\psi_{\eta}^{n}(z,t) = \left(\frac{1}{\cosh\eta}\right)^{(n+1)}
\sum_{k} \left[\frac{(n+k)!}{n!k!}\right]^{1/2}
(\tanh\eta)^{k}\chi_{n+k}(z)\chi_{n}(t) ,
\end{equation}
where $\chi_n(z)$ is the $n$-th excited state oscillator wave function
which takes the familiar form
\begin{equation}
\chi_n (z) = \left[\frac{1}{\sqrt{\pi}2^n n!}\right]^{1/2}
       H_n(z) \exp{\left(\frac{-z^2}{2}\right)} .
\end{equation}
If the hadron is at rest, there are no time-like oscillations,
but for a moving hadron there are.  This is the way in which the
space and time variable mix covariantly and also provides a
resolution of the space-time asymmetry pointed out by Dirac in
his 1927 paper~\cite{dir27}.
\subsection{Probability Interpretations}\label{proba}
The Lorentz-covariant solution given in Eq.(\ref{sol44}) is totally
self-consistent with the quantum probability interpretation.
However, this requires an interpretation of oscillator excitations
along the time-separation coordinate $t$\cite{knp86,kn77par}.
We shall study this in terms of two harmonic oscillators.
Let us start with a two-oscillator system with the Hamiltonian of
the form
\begin{equation}\label{hamil+}
H_{+} = \frac{1}{2} \left\{\left[-\left(\frac{\partial}
{\partial x_{1}}\right)^2 + x_{1}^2 \right]
+ \left[-\left(\frac{\partial}{\partial x_{2}}\right)^2 +
x_{2}^2\right]\right\} ,
\end{equation}
and the equation
\begin{equation}\label{diff88}
H_{+} \psi\left(x_{1},x_{2}\right) = \left(n_{1} + n_{2} + 1\right)
\psi\left(x_{1},x_{2}\right) .
\end{equation}
This is the Schr\"odinger equation for the two-dimensional harmonic
oscillator.  The differential equation is separable in the $x_1$ and
$x_2$ variables, and the wave function can be written
\begin{equation}
\psi\left(x_{1},x_{2}\right) = \chi_{n_1}\left(x_{1}\right)
 \chi_{n_2}\left(x_{2}\right) ,
\end{equation}
where $\chi_n(z)$ is the $n^{th}$ excited-state oscillator wave function
which takes the form
\begin{equation}\label{chi11}
\chi_n (x) = \left[\frac{1}{\sqrt{\pi}2^n n!}\right]^{1/2}
       H_n(x) \exp{\left(\frac{-x^2}{2}\right)} .
\end{equation}
Thus
\begin{equation} \label{cwf66}
\psi\left(x_{1},x_{2}\right) =
 \left[\frac{1}{\pi 2^{(n_1 + n_2)} n_1! n_2!}\right]^{1/2}
 H_{n_1}\left(x_1\right) H_{n_2}\left(x_2\right)
  \exp{\left[- \frac{1}{2}\left(x_1^2 + x_2^2 \right)\right]} .
\end{equation}
\par
If the system is in ground state with $n_{1} = n_{2} = 0$, the above
wave function becomes
\begin{equation}\label{cwf99}
\psi\left(x_{1},x_{2}\right) = \left[\frac{1}{\pi}\right]^{1/2}
 \exp{\left[- \frac{1}{2}\left(x_1^2 + x_2^2 \right)\right]} .
\end{equation}
If only the $x_2$ coordinate is in its ground state, the wave function
(with $n = n_1$) becomes
\begin{equation}\label{cwf88}
\psi\left(x_{1},x_{2}\right) =
        \left[\frac{1}{\pi 2^n n!}\right]^{1/2}
       H_{n} \left(x_1\right)
       \exp{\left[- \frac{1}{2}\left(x_1^2 + x_2^2 \right)\right]} .
\end{equation}
\par
If we introduce the normal coordinate system with
\begin{equation}\label{normal11}
y_1 = \frac{1}{\sqrt{2}} \left( x_{1} + x_{2}\right), \qquad
y_2 = \frac{1}{\sqrt{2}} \left( x_{1} - x_{2}\right) ,
\end{equation}
and set
\begin{equation}
y_{1} \rightarrow e^{\eta} y_1,  \qquad
y_{2} \rightarrow e^{-\eta} y_1,
\end{equation}
we can derive the equation\cite{knp86,knp91}
\begin{equation}\label{cwf331}
\psi_{\eta}^{n}\left(x_1,x_2\right) =
 \left(\frac{1}{\cosh\eta}\right)^{(n+1)}
\sum_{k} \left[\frac{(n+k)!}{n!k!}\right]^{1/2}
(\tanh\eta)^{k}\chi_{n+k}\left(x_1\right)\chi_{n}\left(x_2\right) .
\end{equation}
This wave function is a linear combination of the eigen
functions which satisfies the
eigenvalue equation with the Hamiltonian $H_{-}$, where
\begin{equation}\label{hamil-}
H_{-} = \frac{1}{2} \left\{\left[-\left(\frac{\partial}{\partial x_{1}}\right)^2
+ x_{1}^2 \right]
- \left[-\left(\frac{\partial}{\partial x_{2}}\right)^2 +
x_{2}^2\right]\right\} .
\end{equation}
Then
\begin{equation}
H_{-} \chi_{n}\left(x_1\right)\chi_{m}\left(x_2\right)
= (n - m) \chi_{n}\left(x_1\right)\chi_{m}\left(x_2\right) .
\end{equation}
If the $x_2$ coordinate is in its ground state,
\begin{equation}\label{diff99}
H_{-} \psi\left(x_{1},x_{2}\right) = n \psi\left(x_{1},x_{2}\right) .
\end{equation}
If we replace the notations $x_{1}$ and $x_{2}$ by $z$ and $t$
respectively, this Hamiltonian becomes that of Eq.(\ref{diff44}).

\subsection{Time-separation variable}
We now understand the covariant harmonic oscillators in terms of the
two coupled oscillators.  In the case of the coupled oscillators,
coordinates for both oscillators are well defined and carry their
physical interpretation.  However, in the covariant oscillators,
the time-separation variable is still problematic.

\par
This variable exists according to Einstein, and the differential
equation of Eq.(\ref{diff33}) is Lorentz-invariant because of this
time-separation variable.  Yet, its role has not been defined in
the present form of quantum mechanics. On the other hand, it is
possible to explain this variable in terms of Feynman's rest of the
universe~\cite{fey72,kiwi90pl,hkn99ajp}.  The failure to observe
this variable causes an increase in entropy and an additon of
statistical uncertainty to the system.

\section{Lorentz-covariant Quark Model}\label{quarkmo}
We started this paper with the Lorentz-invariant differential
equation of Eq.(\ref{diff11}).  This phenomenological equation
can explain the hadron spectra based on Regge trajectories and
hadronic transition rates.
\par
If we separate this equation using hadronic and quark
variables, the equation can describe a
hadron with a definite value for its four-momentum and its internal
angular momentum.  Furthermore, the hadronic mass is determined
by the internal dynamics of the quarks.
Our next question is whether these wave functions, particularly
their Lorentz covariance properties,  are consistent with what we
observe in the real world.
\par
The number of quarks inside a static proton is three, while Feynman
observed that in a rapidly moving proton the number of partons
appears to be infinite~\cite{fey69b,knp86}.   The question then is
how the proton looking like a bound state of quarks to one observer
can appear different to an observer in a different Lorentz frame?
Feynman made the following systematic observations:

\begin{itemize}
\item[a.]  The picture is valid only for hadrons moving with
 velocity close to that of light.
\item[b.]  The interaction time between the quarks becomes dilated,
 and partons behave as free independent particles.
\item[c.]  The momentum distribution of partons becomes widespread as
 the hadron moves fast.
\item[d.]  The number of partons seems to be infinite or much larger
 than that of quarks.
\end{itemize}

Because the hadron is believed to be a bound state of two
or three quarks, each of the above phenomena appears as a paradox,
particularly b) and c) together.  How can a free particle have a
wide-spread momentum distribution?  We have addressed this question
extensively in the literature, and concluded that Gell-Mann's quark
model and Feynman's parton model are two different manifestations of
the same Lorentz-covariant
quantity~\cite{knp86,kn77par,hussar81,kim89,kn05job}.
\par
As for experimental observations of hadronic wave functions,
it was noted by Hofstadter and McAllister in 1955 that the
proton is not a point particle but has a space-time
extension \cite{hofsta55}.  This discovery led to the study
of electromagnetic form factor of the proton.
As early as in 1970, Fujimura {\em et al.} calculated the
electromagnetic form factor of the proton using the wave functions
given in this paper and obtained the so-called ``dipole'' cut-off
of the form factor~\cite{fuji70}.
\par
In our 1973 paper~\cite{kn73}, we attempted to explain the
covariant oscillator wave function in terms of the coherence
between the incoming signal and the width of the contracted wave
function.  This aspect was explained in the overlap of the
energy-momentum wave function in our book~\cite{knp86}.  Without
this coherence, the form factor
could decrease exponentially for increasing
$(momentum~transfer)^2$.  With this coherence,
the decrease is slower and as shown experimentally, inversely
proportional to the
$(momentum~transfer)^2$.
\section*{Conclusions}
The focal point of this paper is the Lorentz-invariant
differential equation of Feynman {\em et al.} given in
Sec.~\ref{feynman}.  This equation can be separated into the
Klein-Gordon equation for the hadron and a harmonic-oscillator
equation for the quarks inside the hadron.
\par
From the solutions of this equation, it is possible to
construct a representation of Wigner's little group for massive
particles.  These solutions are consistent with both quantum
mechanics and special relativity.  Those oscillator solutions
also explain Dirac's efforts summarized in Fig.~\ref{dirkn}.
In this way, we have combined the contributions made by Wigner, Dirac
and Feynman to make quantum mechanics of bound states consistent
with relativity.
\par
We have also compared the covariant formalism with what we
observe in high-energy physics, specifically the proton
form factor and Feynman's parton picture.

\end{document}